\documentclass{article}
\usepackage[utf8]{inputenc}
\usepackage{authblk}
\usepackage{setspace}
\usepackage[margin=1.25in]{geometry}
\usepackage{graphicx}
\graphicspath{ {./figures/} }
\usepackage{subcaption}
\usepackage{amsmath}

\usepackage[style=ieee, 
citestyle=numeric-comp,
sorting=none]{biblatex}
\title{Degenerate skyrmionic states in synthetic antiferromagnets}

\author{Mona Bhukta}

    \author{Braj Bhusan Singh}
\author[2]{Sougata Mallick}
\author[2]{Stanislas Rohart}
\author[1,*]{Subhankar Bedanta}

\affil[1]{Laboratory for Nanomagnetism and Magnetic Materials, School of Physical Sciences, National Institute of Science Education and Research (NISER), An OCC of Homi Bhabha National Institute (HBNI), Jatni 752050, India}
\affil[2]{Laboratoire de Physique des Solides, Université Paris-Saclay, CNRS UMR 8502, F-91405 Orsay Cedex, France}
\affil[*]{Corresponding author. sbedanta@niser.ac.in}

\date{}

\begin{document}

\maketitle

\begin{abstract}
Topological magnetic textures, characterized by integer topological charge $S$, are potential candidates in future magnetic logic and memory devices, due to their smaller size and expected low threshold current density for their motion. An essential requirement to stabilize them is the  Dzyaloshinskii-Moriya interaction (DMI) which promotes a particular chirality, leading to a unique value of $S$ in a given material. However, recently coexistence of skyrmions and antiskyrmions, with opposite topological charge, in frustrated ferromagnets has been predicted using $J_1$--$J_2$--$J_3$ classical Heisenberg model, which opens new perspectives, to use the topological charge as an additional degree of freedom. In this work, we propose another approach of using a synthetic antiferromagnetic (SAF) system, where one of the ferromagnetic (FM) layer has isotropic and the other FM layer has anisotropic DMI to promote the existence of skyrmions and antiskyrmions, respectively. A frustrated interaction arises due to the coupling between the magnetic textures in the FM layers, which enables the stabilization and coexistence of 6 novel elliptical topological textures.

\end{abstract}

\section{\label{sec:level1}Introduction}

Ever increasing demand of device miniaturization and
low power consumption has led to shift in the research interest from conventional storage technology towards low
dimensional magnetic solitons, viz. chiral domain walls, magnetic vortices, skrymions \cite{skyrme}, etc.
Skyrmionics has drawn significant research  interest in nanotechnology  because of its potential in future magnetic logic and memory devices due to their topological protection, small size, and the ability to set them into motion with significantly smaller threshold current density \cite{fert2013skyrmions, finocchio2016magnetic}. Skyrmions may be used for the development of the futuristic  neuromorphic and probabilistic based computers \cite{jakob_thermal, neuro,jena_gbel_ma_kumar_saha_mertig_felser_parkin_2020}.
\begin{figure}[h!]
    \centering
    \includegraphics[width = 13cm]{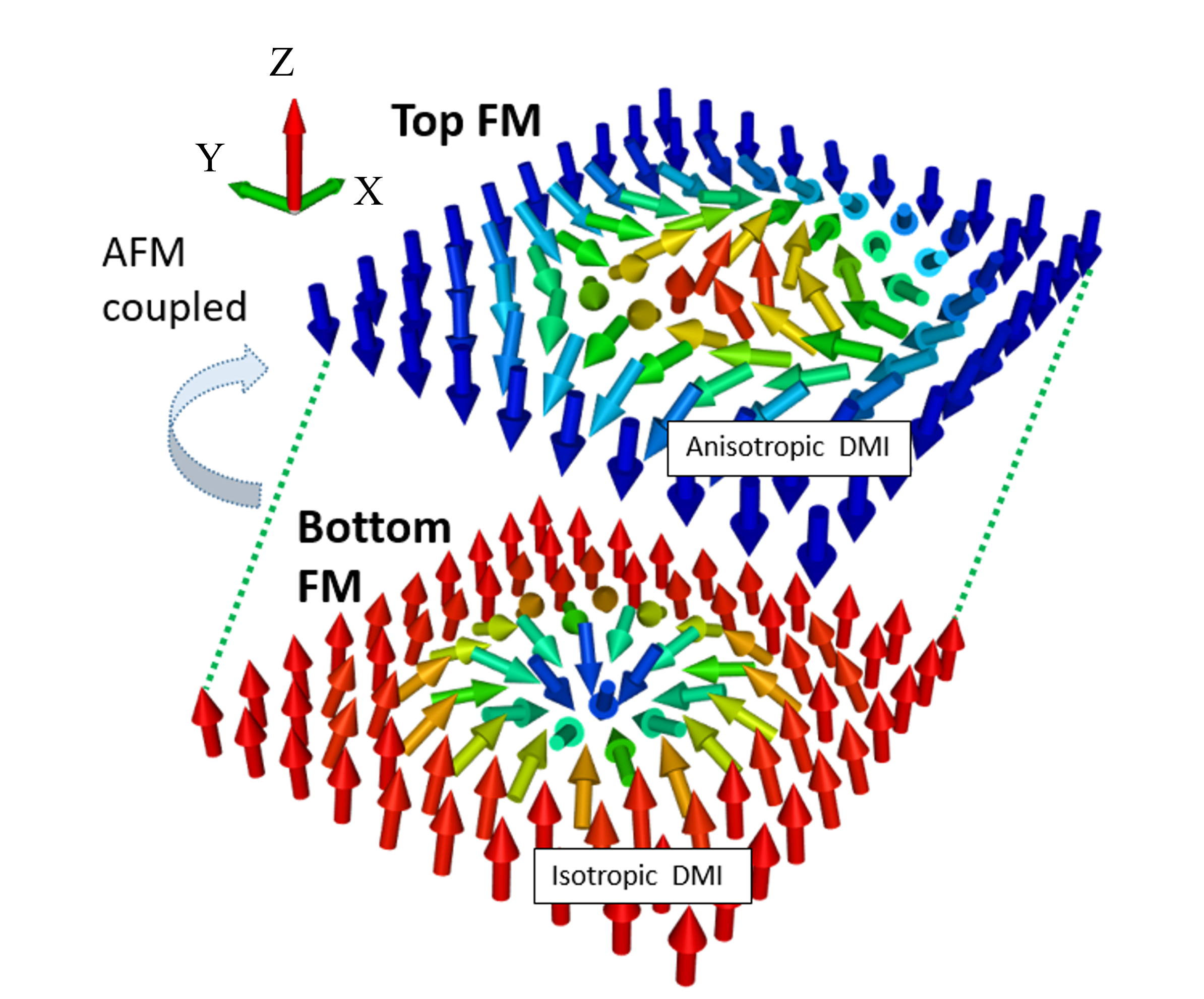}
    \caption{Schemetic of a texture (antiskyrmion-skyrmion pair) in a SAF system where the top and bottom FM layers have anisotropic, and isotropic DMI, respectively.}
    \label{system}
\end{figure}
Topological textures in 2-dimensional media, like skyrmions~\cite{skyrme,bogdanov1989thermodynamically}, can be classified based by a winding number ($W$), an integer that counts the number of times the local magnetization rotates along a closed circuit around the texture core. ~\cite{nagaosa2013topological,chen_2017}. Considering the texture core polarity $p$, the topological charge $S$ is defined as $S = pW$~\cite{nagaosa2013topological}. It should be noted that $W = 1$ represents a skyrmion , whereas it is -1 for an antiskyrmion (see fig 1)
 \cite{camosi2018micromagnetics}. These textures are stabilized by various interactions such as (i) Dzyaloshinskii-Moriya interaction (DMI) \cite{dzyaloshinsky1958thermodynamic, moriya1960anisotropic,fert1981}, (ii) Heisenberg (frustrated) exchange \cite{okubo2012multiple} (iii) dipolar coupling \cite{lin1973bubble, garel1982phase} (iv) four spin exchange \cite{heinze2011spontaneous} etc. Further these magnetic textures may have different size and energetically favored configurations~\cite{nagaosa2013topological}.

 Skyrmions and antiskyrmions can be stabilized in a magnetic film in the presence of isotropic and anisotropic DMI,  respectively\cite{binz_vishwanath_aji_2006,muhlbauer2009skyrmion, yu_onose_kanazawa_park_han_matsui_nagaosa_tokura_2010, yu_kanazawa_onose_kimoto_zhang_ishiwata_matsui_tokura_2010, yu2012skyrmion,camosi2018micromagnetics}. In most thin film systems, the DMI is isotropic due to the polycrystalline nature of the ferromagnetic (FM) films. Such situation leads to formation of skyrmions with a well defined chirality. In order to stabilize antiskyrmion the DMI should have opposite sign along two perpendicular directions of the film plane\cite{bogdanov1989thermodynamically,camosi2018micromagnetics}. To incorporate an anisotropic DMI, low symmetry materials are required with two anisotropy axes, that can be obtained using expitaxial films \cite{bogdanov1989thermodynamically,hoffmann_zimmermann}.
Exchange frustration, for example in the $J_1$--$J_2$--$J_3$ classical Heisenberg model (with $J_1J_2<0$), has been shown to be a solution to enable, skyrmions and antiskyrmions as well as high topological charge ($|W|>2$) solutions in the same film\cite{okubo2012multiple}. However, this situation is rather specific to a limited combination of materials and can hardly be generalized.

On the other hand, topology of a skyrmion/antiskyrmion leads to the gyrotropic force, that acts on the moving skyrmion, pointing perpendicular to its velocity\cite{hoffmann2017skyrmion}. This deviates the path of the skyrmion towards the edge of the nanotrack and this phenomenon is referred as the skyrmion Hall effect (SkHE)\cite{jiang2017direct,litzius2017skyrmion}. In a synthetic antiferromagnet (an antiferromagnetically exchange-coupled FM bilayer) nanotrack, this SkHE could be suppressed without affecting the topology of the skyrmion \cite{zhang_xia_zhou_wang_liu_zhao_ezawa_2016, zhang2016antiferromagnetic, gobel2017antiferromagnetic}.

In this work, we have considered a bilayer synthetic antiferromagnet (SAF) with top and bottom layers consisting of anisotropic and isotropic DMI, respectively (schematically represented in fig \ref{system}). A bound state, that consists of two coupled textures  in each layers, necessarily imply a frustration in the interlayer exchange energy or DMI. This mimics the frustrated Heisenberg exchange discussed earlier. We present coexistence of 6 novel elliptical skyrmionic states with different topological charges as a function of the interlayer coupling.



 \section{Methodology }
\begin{figure*}
    \centering
    \includegraphics[width = 15cm]{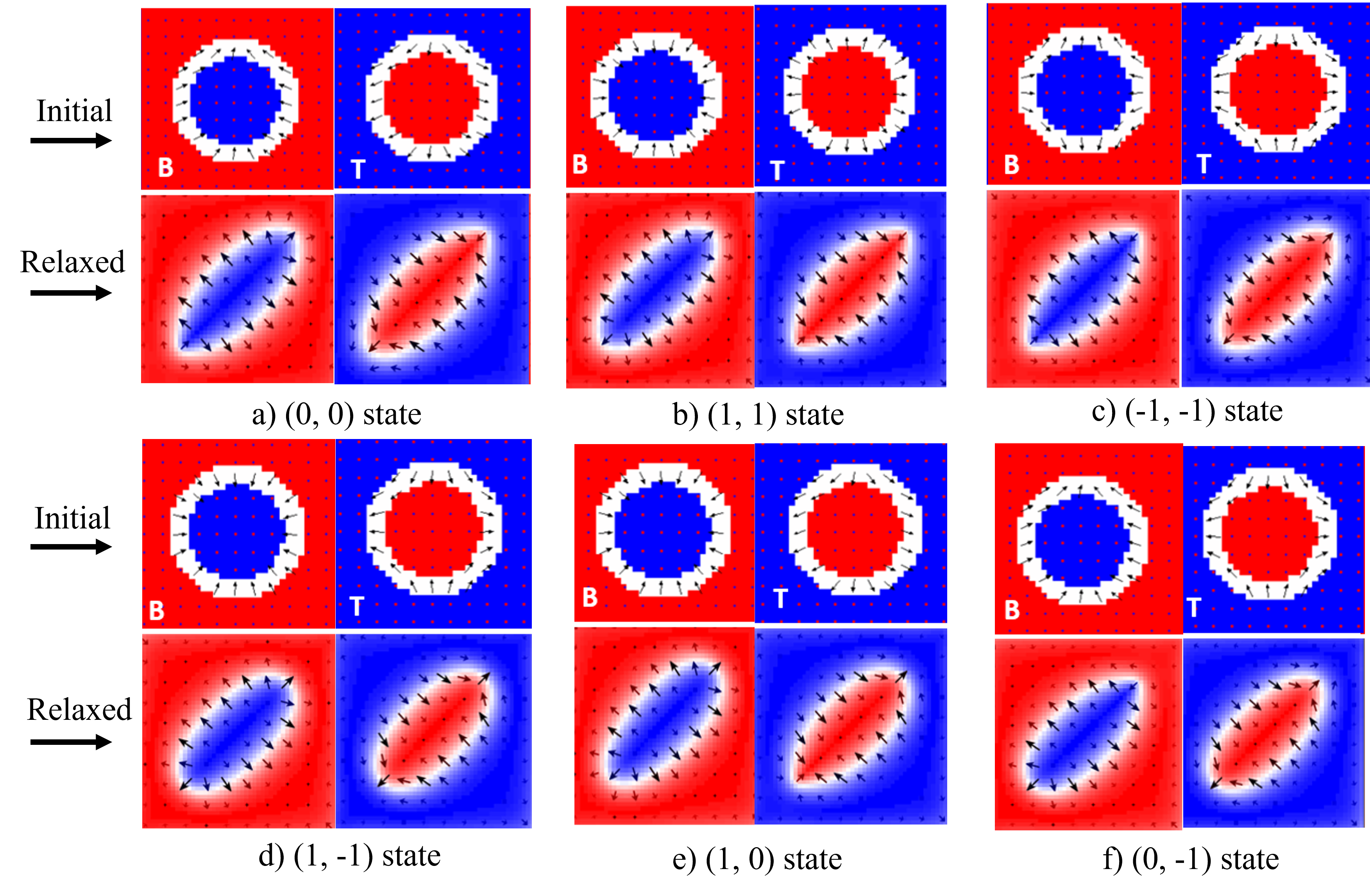}
    \caption{(a-f) Metastable elliptical ground state at intermediate coupling limit ($J_{RKKY}$ = $-8\times 10^{-4}$ $J/m^2$). In each layers the initial and relaxed textures are represented.$W_{bottom}$ and $W_{top}$ are the winding numbers of the bottom and top layers, respectively.
    }
    \label{inter1}
\end{figure*}
We have performed micromagnetic simulations by using the Object-Oriented Micromagnetic Framework (OOMMF) software \cite{donahue}. It solves the time dependent spin dynamics governed by Landau-Lifshitz-Gilbert (LLG) equation. Details about the micromagnetic simulations is discussed in the supplementary information. The system consists of two FM layers antiferromagnetically coupled by Ruderman–Kittel–Kasuya–Yosida (RKKY) coupling. The two layers present the same spontaneous magnetization, isotropic Heisenberg exchange and perpendicular magnetic anisotropy (PMA). In the top layer the DMI constant $D_{iso}$ is isotropic, whereas in the bottom layer the DMI constant has opposite sign along $x$ and $y$ direction, but with the same absolute value $D_{aniso}$. The spins of top and bottom layers
point downward (blue color) and upward (red
color) directions, respectively, as shown in fig \ref{system}.

In this work, we have simulated a $100\times100\times3$~nm$^3$ sample with a $2\times2\times1$~nm$^3$ cell size and open boundary conditions. The 3 nm thickness considered here incorporates 1 nm thick top and bottom magnetic layers and 1 nm thick metallic spacer layer to introduce RKKY coupling. The simulation parameters are spontaneous magnetization $M_S= 1.2\times 10^{6}$~A/m, exchange constant $A=16\times 10^{-12}$~J/m, PMA $K=0.16\times 10^{5}$~J/m$^3$ and Gilbert damping constant $\alpha = 0.1$. In the bottom layer the DMI is isotropic with a strength $D_{iso}=0.0025$~J/m$^2$. In the bottom layer, the DMI is anisotropic with a stength $D_{aniso}=0.0025$~J/m$^2$ along $x$ and $-D_{aniso}$ along $y$.  These parameters are similar to the values taken in reference \cite{camosi2018micromagnetics}. We have varied the strength of the RKKY interaction $J_{RKKY}$ from $-8\times 10^{-3}$ to   $-1\times 10^{-5}$~J/m$^{2}$. Magnetic configurations have been relaxed solving the LLG equation up to a stopping criterion $|\frac{dm}{dt}| \leq 0.001$ degree/ns.


\section{Results and discussion}

\begin{figure}[h!]
    \centering
    \includegraphics[width = \textwidth]{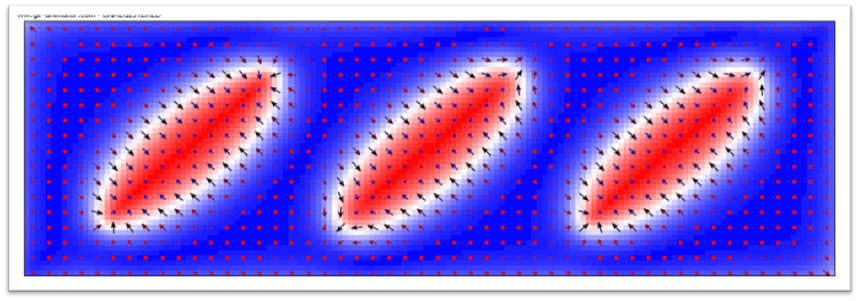}
    \caption{Co-existence of a $W=1$ skyrmion, an antiskyrmion, and $W=0$ skyrmion in the top layer of the SAF at
    $J_{RKKY}=-8\times 10^{-4}$~J/m$^2$.}
    \label{three}
\end{figure}

When the top and bottom layers containing their natural textures (i.e. antiskyrmion and skyrmion, respectively) are antiferromagnetically coupled, the interlayer coupling cannot completely be in energetically favourable condition in all directions. Hence, there will be an energy frustration in the system. We have simulated the proposed model to explore the effect of the strength of the RKKY coupling on the textures by relaxing 

9 different initial textures combining $W = 1$, 0 and -1 in both the layers, We have denoted the states as ($W_{bottom}$, $W_{top}$), where $W_{bottom}$ and $W_{top}$ are the winding numbers of the bottom and top layers, respectively. 

At low coupling strength, from $-8\times 10^{-5}$ to $-1\times 10^{-5}$~Jm$^{-2}$, the relaxed states show no significant change as compared to the system having no inter-layer coupling. Here, only the formation of skyrmion-antiskyrmion pair was obtained in the two layers.

Nevertheless, in an intermediate coupling range, from $-8\times 10^{-4}$ to $-2\times 10^{-4}$~Jm$^{-2}$, new stable textures were obtained.

In this coupling limit, three states: (-1, 1), (0, 1), and (-1, 0), exhibit to texture collapse after relaxation. These states are neither in an energetically favorable condition in terms of the interlayer coupling nor do they satisfy the symmetry of the DMI in their respective layers. The other 6 meta-stable states ((0, 0), (0, -1), (1, 1), (1, -1), (1, 0), and (-1, -1)) are shown in the fig \ref{inter1}. Three states among them ((0, 0), (1, 1), and (-1, -1)) present the same texture in each layer (fig.~\ref{inter1}(a-c)), which satisfy the interlayer coupling, however, costs some DMI energy. It should be noted that surprisingly, a (0, 0) texture is stable, which satisfies none of the DMI symmetries, however is stable solely due to the interlayer coupling, similarly to exchange frustrated systems~\cite{rozsa2017formation}. Fig.~\ref{inter1}(d-f) show the three remaining configuration where textures from the two layers with different winding numbers are coupled. Among them, the $(1,-1)$ texture naturally satisfies the energy requirement of DMI in each layers, however, it costs a significant amount of interlayer coupling energy which is discussed later. Two remaining textures ((1, 0) and (0, -1)) exhibit presence of $W=0$ skyrmion in one of the layers. Along with the existence of  $W = 0$ skyrmion, skyrmion and antiskyrmion in a single  film have also been stabilized. Figure \ref{three} shows the coexistance of a skyrmion, an antiskyrmion and a $W=0$ skyrmion in the top layer of a SAF in where $J_{RKKY} =  -8 \times 10^{-4}$~J/m$^2$.

 \begin{figure}[h!]
    \centering
    \includegraphics[width = 15cm]{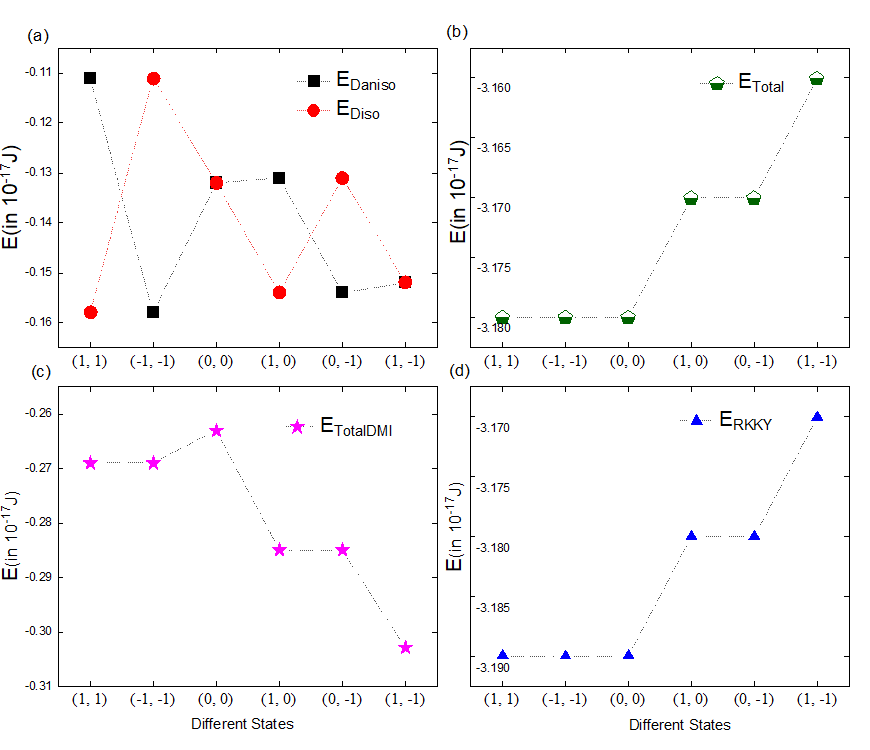}
    \caption{Energy distribution of six metastable states ((0, 0), (1, 1), (-1,-1), (1,0), (0,-1) and (1,-1)) as a function of (a) $E_{Diso}$, $E_{Daniso}$; (b) $E_{Total}$; (c) $E_{TotalDMI}$; (d) $E_{RKKY}$, at $J_{RKKY} = -8 \times 10^{-4}$~J/m$^2$.}
    \label{inter2}
\end{figure}



The textures obtained after relaxation are elliptical in shape in the presence of interlayer coupling, in contrary to the circular textures in its absence. This can be qualitatively explained since the DMI strength in both layers are the same along the $x$ axis, but have opposite signs along the $y$ axis. Therefore, an effective DMI is induced in the system due to the interlayer coupling which can be estimated by adding the individual DMI contributions of the two layers. It is maximum along the $x$ axis and minimum along the $y$ axis, leading to  an elliptical distortion of the structure. It is noted that, the formation of similar elliptical skyrmions have been observed experimentally in samples with anisotropic DMI \cite{jena_gbel_ma_kumar_saha_mertig_felser_parkin_2020,peng_takagi_koshibae_shibata_arima_nagaosa_tokura_seki_yu_2019}. The emergence of elliptical skyrmions in such systems has been corroborated to either the effect of dipole-dipole interaction \cite{jena_gbel_ma_kumar_saha_mertig_felser_parkin_2020}, or to the consequence of in-plane applied magnetic field \cite{peng_takagi_koshibae_shibata_arima_nagaosa_tokura_seki_yu_2019}. In order to understand the mechanism of formation of elliptical textures in this work, we considered three different cases: (i) finite RKKY coupling, zero dipolar coupling; (ii) finite RKKY coupling, finite dipolar coupling; (iii) zero RKKY coupling, finite dipolar coupling. We observe that the occurrence of the elliptical shape depends solely on the presence of a finite RKKY coupling irrespective of the dipolar coupling.
\begin{figure}[htb]
    \centering
    \includegraphics[width = 13.0cm]{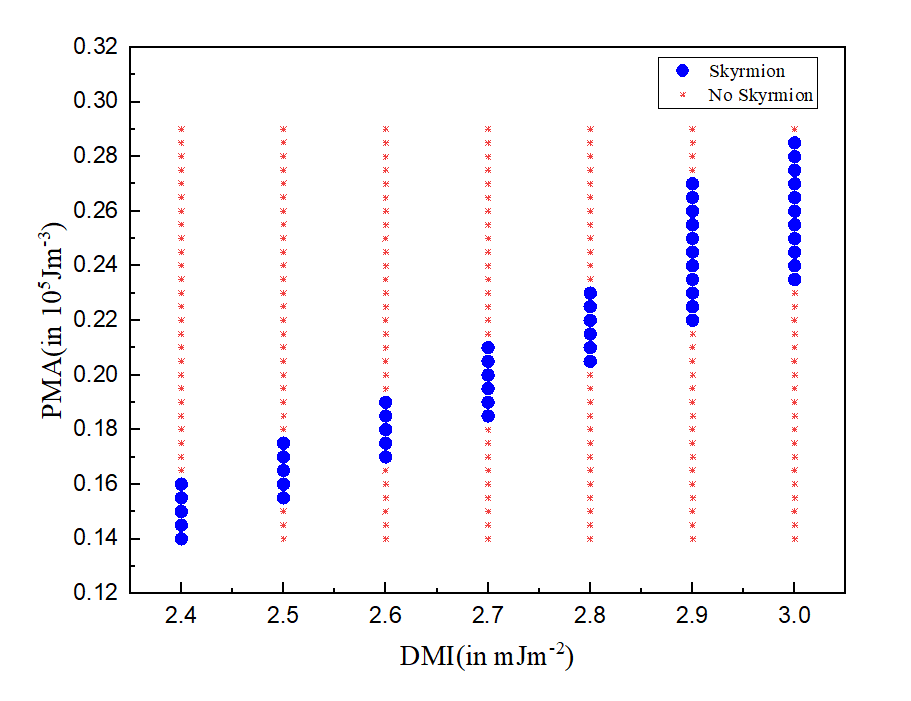}
    \caption{DMI and K phase diagram  of skyrmionic textures at $J_{RKKY}$ = $-8\times 10^{-4}$ $J/m^2$. The blue circles and red cross represent the combination of DMI and K  where the textures could be stabilized and not stabilized, respectively.}
    \label{dk}
\end{figure}

Different contribution of the DMI energies i.e  $E_{Diso}$, $E_{Daniso}$, and $E_{Dtotal}$ for all the six energetically favorable states have been extracted from the simulations and are presented in fig \ref{inter2}(a-c). We observe a three fold degeneracy in total energy ($E_{Total}$) among the (0, 0), (1, 1) and (-1, -1) states. Further, a two fold degeneracy is also observed between (0, -1) and (1, 0) states. Fig. \ref{inter2}(d) shows the degenerate energy levels arising from the sole contribution of RKKY coupling. 

\begin{figure}[htb]
    \centering
    \includegraphics[width = 12cm]{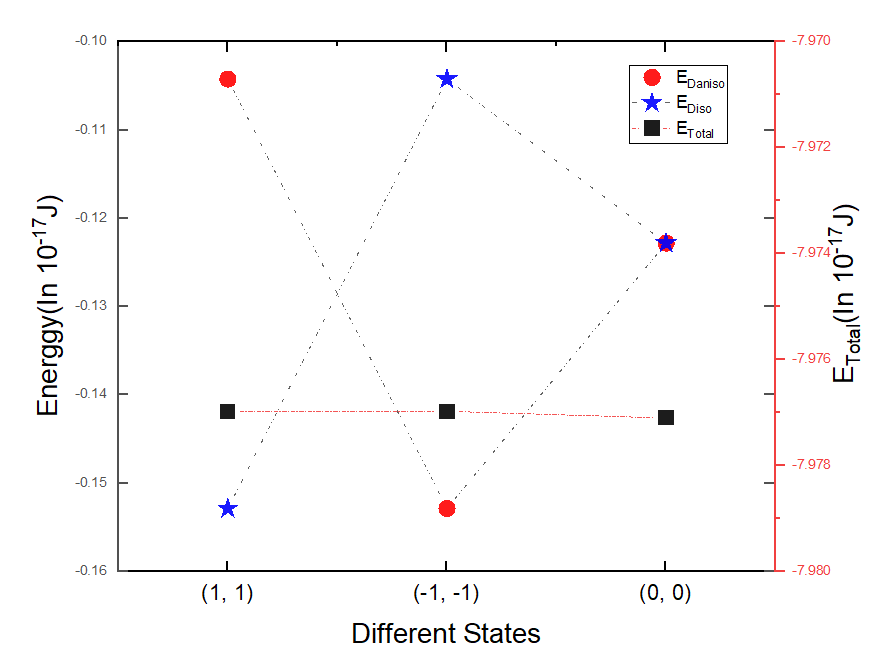}
    \caption{Energy distribution of three states (0, 0), (1, 1) and (-1, -1) at $J_{RKKY}$ = $-2\times 10^{-3}$ $J/m^2$. }
    \label{inter3}
\end{figure}
In order to check the validity of our discussion in the intermediate coupling limit, we have performed a systematic scan of DMI and K values where stable textures can be obtained. Fig \ref{dk} shows the acceptable range of DMI and K values for $J_{RKKY}$ = $-8\times 10^{-4}$ $J/m^2$. The blue circles and red crosses represent the set of DMI and K where the textures are stable and not-stable, respectively. In our work, we have chosen a value of DMI and K which can stabilize the textures and are experimentally realized in real samples.

In a strong coupling limit ($J_{RKKY}$ from $-8\times 10^{-3}$ to $-2\times 10^{-3}$~Jm$^{-2}$), we have also observed the signature of induced frustration due to the presence of isotropic and anisotropic DMI in the layers. In contrast to intermediate coupling range, only three possible meta-stable states, (0,0), (1,1) and (-1, -1), are observed. Here the energy requirement for the interlayer coupling energy is completely satisfied, however, for the DMI it is not satisfied. $E_{Total}$, $E_{Diso}$, and $E_{Daniso}$ of all the three energetically favourable states are presented in fig \ref{inter3}. It is observed from the $E_{Total}$ that these states are degenerate.

It is well known that the SkHE is suppressed in SAF systems. In this context, we have studied the dynamics of the (0, 0) state under spin transfer toque. We observed a maximum velocity of 1700 m/s at a current density of $8\times10^{12}$ A/m$^2$ with zero SkHE(refer to supplementary figure S3). The velocity of the skyrmions exhibit a linear behavior with the applied current density, indicating the  flow regime. The details can be found in the supplementary information.

\section{conclusion}

We have successfully modelled a SAF thin
film system in which equivocation in DMI interaction between two antiferromagnetically coupled layers leads to
spin frustration in both the FM layers. We have studied three different: (i) low (ii) intermediate and (iii)
strong RKKY coupling limits. When the RKKY coupling
strength is very low, no sign of frustration in the system
has been observed. However, at intermediate and high coupling, we obtained textures with energy frustration. The dynamics of $W = 0$ skyrmion (0, 0) reveals
zero SkHE. Our work shows that frustration can be crucial in determining the different degenerate skyrmionic
states. Future work should be focused to replicate these
results experimentally. In this context, one may consider
the top layer to be Fe/W(110) thin film which could give
rise to anisotropic DMI, that is necessary for stabilization
of antiskyrmion \cite{hoffmann2017antiskyrmions}. Further, the effect of Gilbert damping on current induced motion of the skyrmionic states needs to be elucidated\cite{akosa_miron_gaudin_manchon_2016, akosa_ndiaye_manchon_2017, yuan_hals_liu_starikov_brataas_kelly_2014}. Our results show that the simultaneous presence of skyrmions and antiskyrmions in a SAF could give an extra degree of freedom for developing novel devices in the context of nanotechnology. 

\section{acknowledgments}

The authors acknowledge department of atomic energy (DAE), Govt. of India and the Indo-French collaborative project supported by CEFIPRA (IFC/5808-1/2017) for providing the research funding. BBS acknowledges the DST INSPIRE faculty fellowship.SR acknowledges Agence National de la Recherche (Contract No. ANR-17-CE24-0025 TopSky).

\printbibliography

\end{document}